\documentclass[runningheads]{svmult}

\usepackage{makeidx}   
\usepackage{graphicx}  
\usepackage{subeqnar}  
\usepackage{multicol}  
\usepackage{physprbb}  
\makeindex             

\usepackage{psfig}
\newcommand{\sivi}{\hbox{[Si\,{\sc vi}]}}
\newcommand{\arcsec}{\hbox{$^{\prime\prime}$}}
\def\spose#1{\hbox to 0pt{#1\hss}}
\def\gtrsim{\mathrel{\spose{\lower 3pt\hbox{$\mathchar"218$}}
     \raise 2.0pt\hbox{$\mathchar"13E$}}}

\begin{document}
\title*{Investigating ULIRGs in the Near-Infrared:\protect\newline Imaging \& Spectroscopy}
\toctitle{Investigating ULIRGs in the Near-Infrared:
\protect\newline Imaging \& Spectroscopy}
%
%
\titlerunning{Investigating ULIRGs in the Near-Infrared}
%
\author{R. Davies\inst{1}
\and A. Burston\inst{2}
\and M. Ward\inst{2}}
\authorrunning{Davies, Burston, \& Ward}
%
%
\institute{Max-Planck-Institut f\"ur extraterrestrische Physik, 85741 Garching, Germany
\and X-ray Astronomy Group, University of Leicester, Leicester, LE1 7RH, UK}

\maketitle              

\begin{abstract}
We present imaging and spectroscopic observations of 2 nearby
($z<0.1$) ULIRGs from a larger sample, and address the question of
whether the JHK continuum colours and slope might be effective probes
of the nuclear region in searches for AGN.
Certainly there is evidence for significant quantities of hot dust
emission at temperatures $\gtrsim$1000\,K; 
but it may be that rather than pointing to an AGN, this instead tells us
more about the environment and evolution of the star formation.
\end{abstract}

\section{Introduction}

Studies of ULIRGs at different wavelengths face a variety of challenges.
These include attenuation at shorter wavelengths (soft X-ray to
optical) due to high extinction, and poor spatial resolution at longer
wavelengths (mid- to far-infrared). 
A good compromise can be found in the near-infrared.

Although there have been a number of JHK imaging studies of ULIRGs
\cite{san88,car90}, colours have tended to be examined in large
apertures for which dilution from extended emission can be significant.
We present data on 2 objects from a larger sample, with a resolution
$<$0.5\arcsec\ allowing us to probe much closer into the nucleus.
Furthermore, combining imaging data with spectroscopy provides much
tighter constraints on the origin of the continuum, which cannot be
achieved with K-band spectra alone.

\section{IRAS\,23365\,+3604}

\begin{figure}[h]
\centerline{\psfig{file=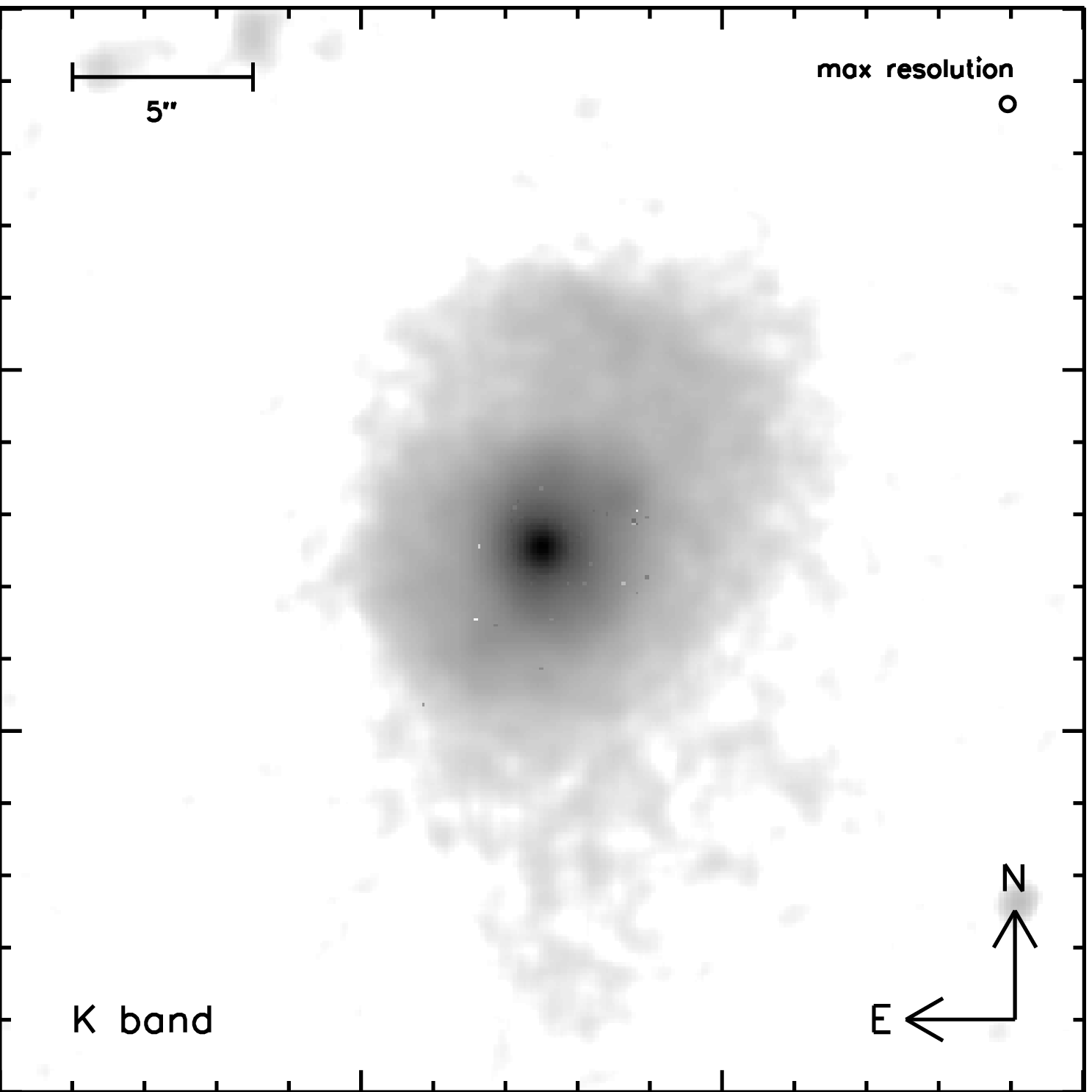,height=5.8cm}\psfig{file=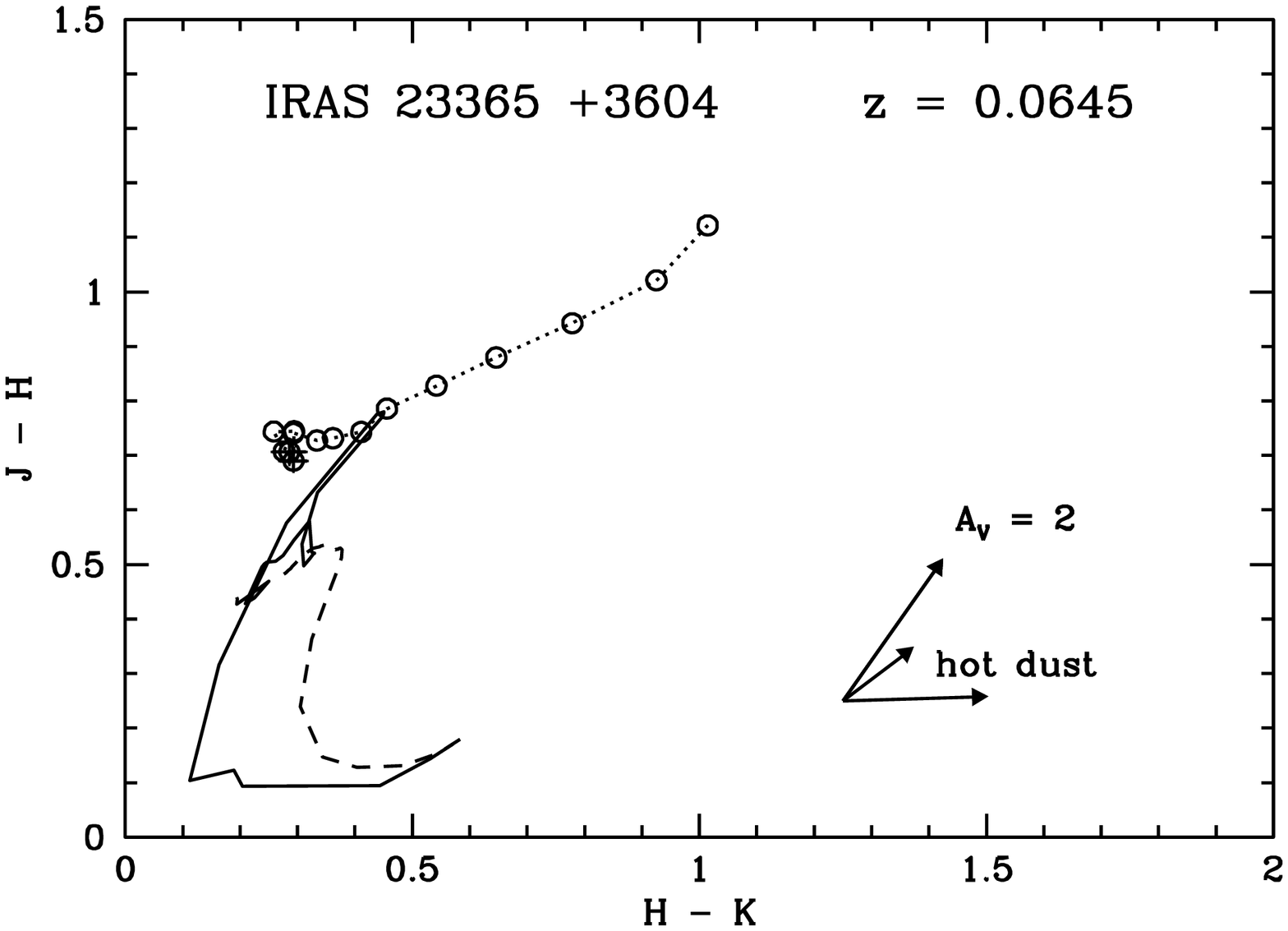,height=5.8cm}}
\caption{Left: K-band image of IRAS\,2336 with a resolution of
0.5\arcsec.
Right: JHK colour diagram showing the colours in annuli of increasing
radius; the nucleus is reddest. Also shown are colours of
instantaneous (solid line) and continuous (dashed line) star formation
from Starburst~99 \cite{lei99}, and the effects of extinction and dust
emission at 500\,K and 1500\,K.}
\label{fig:2336im}
\end{figure}

\begin{figure}[h]
\centerline{\psfig{file=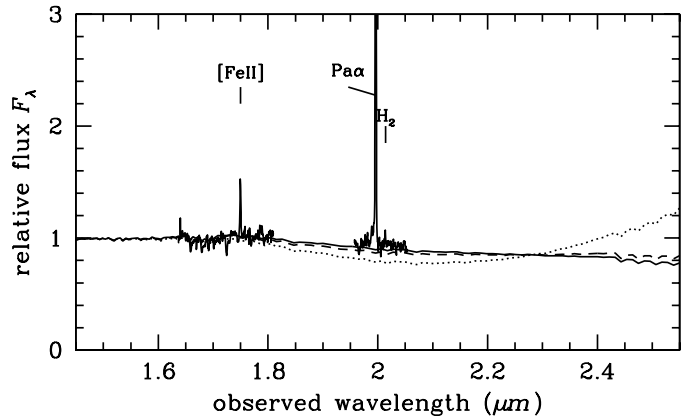,height=5.8cm}}
\caption{Spectrum of IRAS\,2336 extracted in a
0.58\arcsec$\times$0.58\arcsec\ aperture, kindly provided by R.~Genzel
\cite{gen01}.
Overplotted are the 3 models (at low spectral resolution)
derived from the JHK colours above for
dust emission at 1500\,K (solid line), 1000\,K (dashed line), and
500\,K (dotted line).
These show that dust emission below $\sim$1000\,K has a 
characteristic shape which is inconsistent with the data.}
\label{fig:2336spec2}
\end{figure}

\begin{figure}[h]
\centerline{\psfig{file=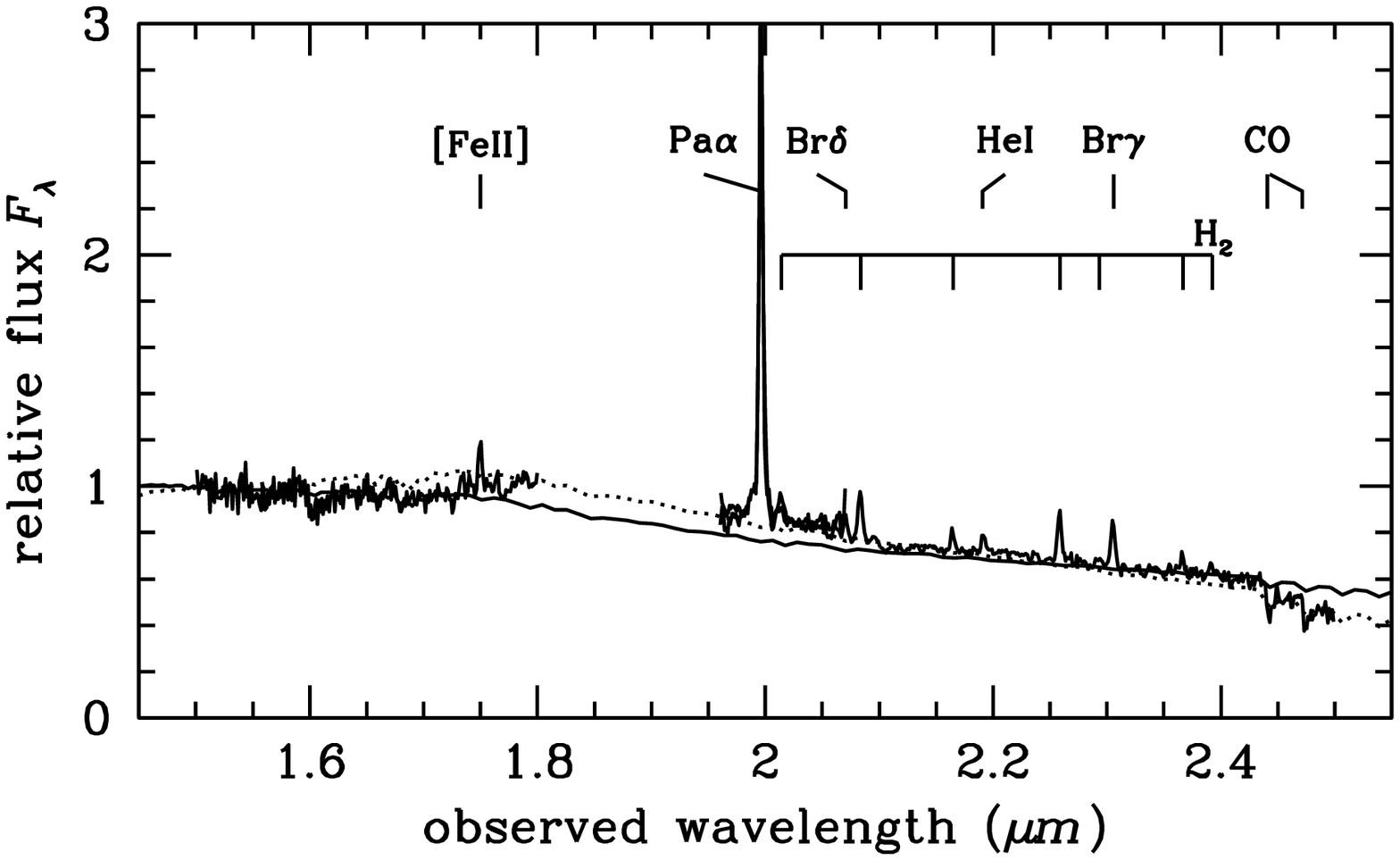,height=5.8cm}}
\caption{Spectrum of IRAS\,2336 extracted in a
0.6\arcsec$\times$3.0\arcsec\ aperture, from \cite{bur01}.
Overplotted are models derived from the JHK colours in a 3\arcsec\
aperture for:
solid line --  a reddened 100\,Myr starburst and hot dust emission at
1500\,K.
dotted line -- a highly reddened starburst with no extra dust emission.}
\label{fig:2336spec1}
\end{figure}

Fig~\ref{fig:2336im} shows the K-band image of IRAS\,2336, which
displays signs of interaction although with no obvious nearby
candidates, at a resolution of 0.40\arcsec. 
The colours in incremental annuli out to 5\arcsec\ are drawn in the
JH-HK colour diagram.
The nucleus (classified as a LINER \cite{vei95} or composite 
\cite{baa98}) is reddest;
colours at larger radii become bluer until beyond a radius of
$\sim$1.5\arcsec\ they stabilise.
For comparison the diagram shows at the redshift of the galaxy, the colours of
stellar populations from \cite{lei99} for instantaneous and
continuous star formation, as well as the effect of
extinction and hot dust.

We have fitted models to the nuclear colours (in a 1\arcsec\
aperture), consisting of a reddened 100\,Myr old stellar population
and a hot dust component.
Except in the extreme case of a very young starburst $<$2.5\,Myr old
-- which is ruled out by the Br$\gamma$ equivalent width --
the nuclear colour cannot be reproduced by reddening stellar light.
Therefore including hot dust (which contributes a fraction $f_{\rm Kd}$ of
the K-band emission) is mandatory.
But since its temperature cannot be constrained from JHK colour data alone,
we consider three particular cases:
$T_{\rm D}$=1500\,K is close to the sublimation temperature of grains
is an upper limit, 
1000\,K is intermediate,
while 500\,K represents the coolest dust that can still
significantly affect the K-band.
With these we find that for 
$T_{\rm D}$=1500\,K, $f_{\rm Kd}$=0.59 and $A_{\rm V}$=2.85; 
for
$T_{\rm D}$=1000\,K, $f_{\rm Kd}$=0.49 and $A_{\rm V}$=3.77; 
and for 
$T_{\rm D}$=500\,K, $f_{\rm Kd}$=0.50 and $A_{\rm V}$=4.36.

The spectrum associated with each of these models is overplotted on a
spectrum of the galaxy nucleus \cite{gen01} in Fig~\ref{fig:2336spec2}.
They show that the model matches the spectrum extremely well, and
constrains the temperature of the dust emssion to be $\gtrsim$1000\,K.
Another spectrum of IRAS\,2336 in Fig~\ref{fig:2336spec1}, discussed
in more detail in 
\cite{bur01}, was extracted in a larger aperture and covers a longer
wavelength range.
The line emission shows none of the typical signs of an AGN:
the Pa$\alpha$ and Br$\gamma$ lines are narrow
(300--500\,km\,s$^{-1}$), and there is no detectable coronal \sivi.
Here we have overplotted models similar to those above, but derived
from colours extracted in 3\arcsec\ apertures;
the difference from the previous figure emphasises the importance of
using consistent apertures when comparing different data.
The model shown (solid line) uses $T_{\rm D}$=1500\,K and matches the
shape of the continuum quite well.
To evaluate whether a model with no dust emission is feasible, we also
show a pure stellar (and nebula) continuum with $A_{\rm V}$=6 (dotted line).
Although this provides a reasonable match to the spectra, it can be
ruled out because its JH colour is too red by 0.32\,mags.
Thus in order to replicate both the JHK colours and the HK spectral
shape, we find that a contribution to the K-band emission from dust at
$T_{\rm D}$$\gtrsim$1000\,K is required.

\section{IRAS\,20210\,+1121}

\begin{figure}[h]
\centerline{\psfig{file=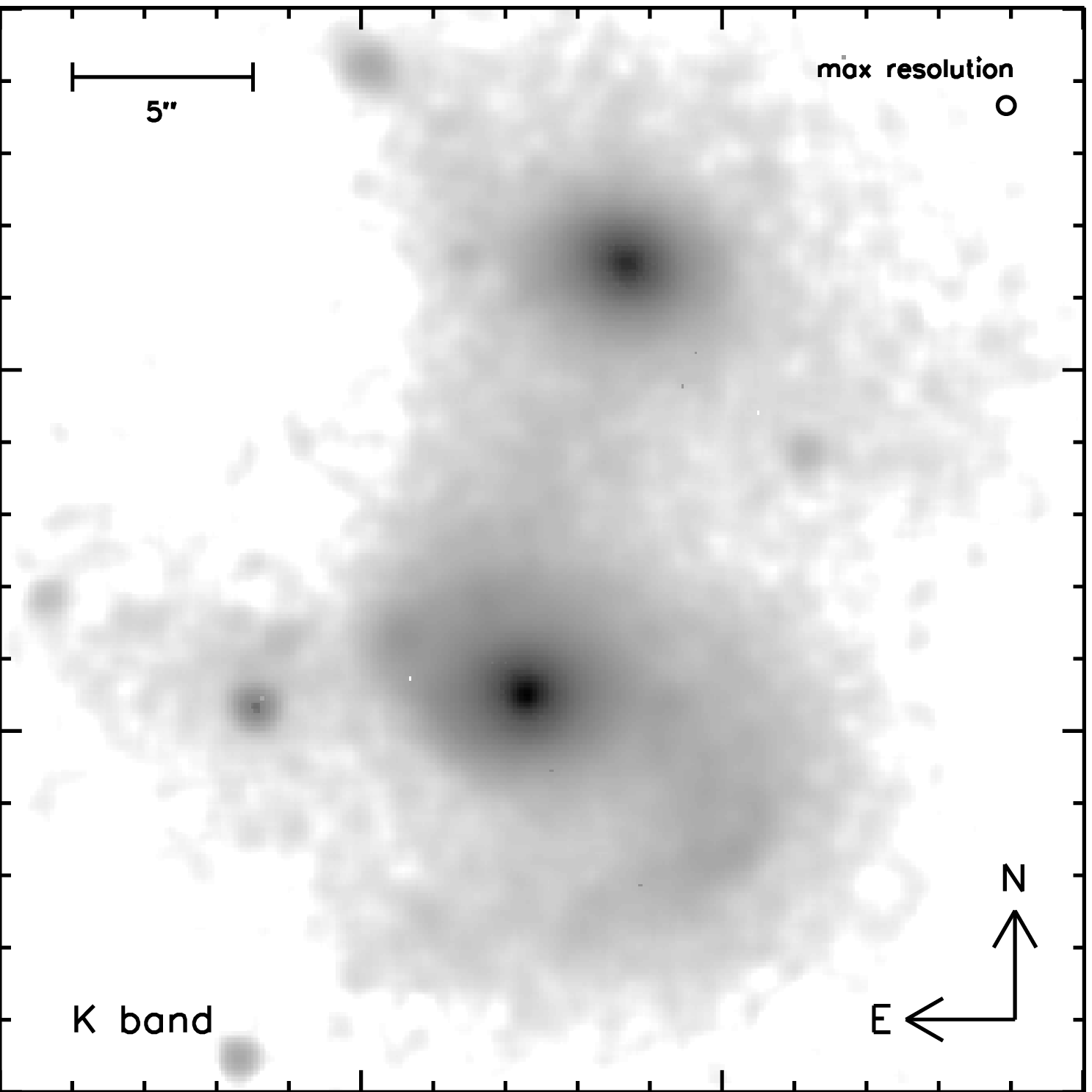,height=5.8cm}\psfig{file=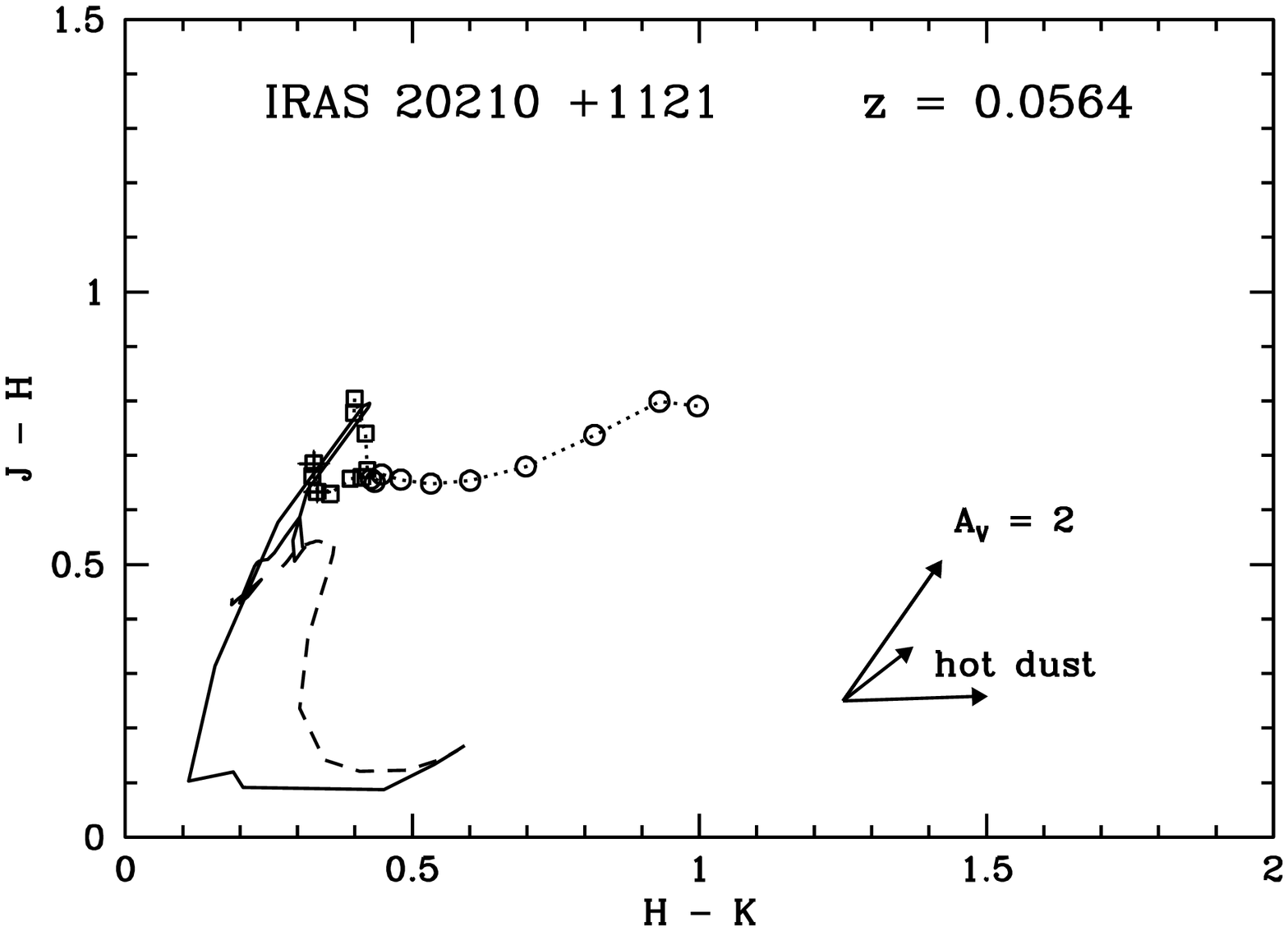,height=5.8cm}}
\caption{Left: K-band image of IRAS\,2021 at a resolution of
0.48\arcsec.
Right: JHK colour diagram of the two nuclei (north, squares; south,
circles) in incremental annuli. Shown are the locus of star formation
models, and the effect of extinction and dust emission at 500\,K and 1500\,K.}
\label{fig:2021im}
\end{figure}

\begin{figure}[h]
\centerline{\psfig{file=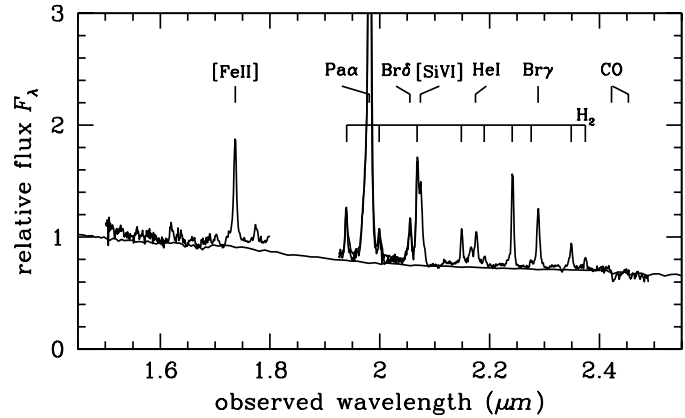,height=5.8cm}}
\caption{Spectrum of IRAS\,2021 extracted in a
0.6\arcsec$\times$3.0\arcsec\ aperture, from \cite{bur01}.
Overplotted is a model derived from the nuclear JHK colours for a 
reddened starburst and hot dust emission at 1500\,K.
It provides a remarkably good fit to the continuum shape.}
\label{fig:2021spec}
\end{figure}

In the K-band image in Fig~\ref{fig:2021im} (resolution 0.48\arcsec),
IRAS\,2021 has two nuclei with a ridge of emission between them.
The JH-HK diagram shows that the colours of annuli
centered on the north nucleus (squares) are consistent with a
stellar population reddened by $A_{\rm V}$$\sim$1.
The circles denote the colours centered on the south nucleus
(classified as a Seyfert~2 \cite{per90,vad98}).
As before, at radii larger than $\sim$1.5\arcsec\ these are typical of a
moderately reddened stellar population, but for the nucleus itself hot
dust is required.
Fitting models to the colours of the south nucleus in a 1\arcsec\
aperture yields:
for $T_{\rm D}$=1500\,K, $f_{\rm Kd}$=0.70 and $A_{\rm V}$=0.04; 
for $T_{\rm D}$=1000\,K, $f_{\rm Kd}$=0.60 and $A_{\rm V}$=1.38;
and for 
$T_{\rm D}$=500\,K, $f_{\rm Kd}$=0.60 and $A_{\rm V}$=2.20.

Although only 2 examples are presented in this contribution, we have
looked at 8 nuclei in 6 ULIRGs, the analysis of which is given in
\cite{dav01}.
For the sample as a whole, varying the age or metallicity of the
stellar population does not affect the results much because, except
for ages less than 10--20\,Myr, the JHK colours hardly change.
It is although -- and because -- the stellar population cannot be
constrained, that the result we find here about significant hot dust
emission being mandatory is robust.
Our results are similar to those found by \cite{sur00}.

Fig~\ref{fig:2021spec} shows spectra for the southern nucleus of
IRAS\,2021 (see \cite{bur01}), which do exhibit typical signs of an
AGN:
the Pa$\alpha$ is broad as expected for a narrow line region
($\sim$700\,km\,s$^{-1}$), and there is easily detectable \sivi\ --
which is stronger than H$_2$~1-0\,S(1) and even twice as strong as
Br$\gamma$.
A model derived from the broadband colours as before (but in a
2\arcsec\ aperture) is over-drawn.
It matches the continuum extremely well through both the H- and
K-bands.
As for IRAS\,2336 we find that not only is a dust component required
but that we can now constrain the temperature to be close to the top
end of the permissible range, $T_{\rm D}$$\gtrsim$1000\,K.

\section{Starburst or AGN ?}

An important question is whether the presence of dust at
$\gtrsim$1000\,K implies an AGN exists.
Typically, heating dust to its sublimation temperature requires a very intense UV radiation field that only occurs within a few parsecs of an AGN, at the inner edge of a putative torus.
Indeed, ISO spectroscopy of classic `template' starburst and AGN galaxies \cite{stu00,lut00} suggests that it is only AGN which have spectral shapes indicative of dust emission at these temperatures.
However, the ISO-SWS 2.4--12\,$\mu$m aperture is large enough to
include a significant fraction of the more extended emission from a
galaxy bulge or disk.
So late-type stars could easily dominate the 2\,$\mu$m emission
observed by ISO and hide any relatively weaker hot dust emission that
might originate only in younger star clusters.

Evidence that stellar processes can produce dust hotter than 1000\,K
arises from 3 reflection nebulae that were observed to have
2--5\,$\mu$m\ continua characterised by a colour temperature of
$\sim$1000\,K \cite{sel83}.
This is explained in terms of stochastic heating of very small
grains of radius 5--10\AA\ \cite{sel84}, with the result that the fraction of the dust mass in such grains is 0.002, and the fraction of stellar luminosity absorbed and re-radiated by them is 0.004 of that absorbed and re-radiated by all grains.
A simple calculation comparing the luminosities for dust emission at 2.2\,$\mu$m\ and 60\,$\mu$m yields ratios of $\sim$0.01, consistent with this model.

A possible problem with applying this model to ULIRGs is that the
reflection nebulae contain non-ionising stars; 
whereas ULIRGs must host significant numbers of ionising stars, from
which the UV radiation could destroy very small grains and
the thermal nebula emission could mask small grain emission. 
However, one plausible hypothesis for the star formation history in
ULIRGs is that individual star clusters (or groups of clusters) have
formed in multiple episodes spread over a timescale of several
100\,Myr -- effectively continuous star formation.
In such a scheme only the youngest clusters, in which star formation occured within the last 10\,Myr, host ionising stars.
Most of the far-infrared luminosity still originates in these clusters through thermal heating of dust grains; 
but the bulk of the starburst population resides in older clusters
without H{\sc ii}~regions, and could have an observable small grain
population.
The 1000\AA\ luminosity of these 10--100\,Myr clusters is an
order of magnitude greater than that at 2\,$\mu$m, so 
there are enough energetic photons to heat the small grains
and hence account for the equal 
contributions from dust and stars to the 2\,$\mu$m emission.
One could also speculate that in the turbulent environment expected in
the nuclear region of a ULIRG, grain destruction through shocks -- via
the merging process as well as the high supernova rate -- cause an over-density of very small grains.
Thus it is certainly possible that in a ULIRG the necessary
prerequisits are met so that stochastic excitation of small grains
becomes an observable phenomenon.

\section{Conclusion}

As part of a larger study, we present an analysis
which combines JHK imaging and HK spectroscopy of 2 ULIRGs, one that
harbours a Seyfert~2 nucleus and one which shows none of the typical
signs of an AGN.
The crucial result is that models of reddened star formation cannot
explain the JHK colours of the nuclei;
additional hot dust emission is needed.
Although the temperature is unconstrained by colours alone, comparison
to the spectra show it is $\gtrsim$1000\,K.
Hot dust emission in the K-band is usually associated with
the inner edge of a torus around an AGN, but we have
presented evidence that in ULIRGs it could also arise through
stochastic heating of very small grains by non-ionising stars.

\section{Acknowledgements}
RID thanks the many people at the meeting, particularly J.~Graham and
M.~Lehnert, who provided helpful input to this work.
Some of the data here were obtained as part of the UKIRT
Service Programme, and we thank all those who were involved.
UKIRT is operated by the JAC on behalf of the U.K. PPARC.

%

\end{document}